\documentclass{article}

\PassOptionsToPackage{numbers, sort&compress}{natbib}

\usepackage[final]{neurips_2021_ml4ps}

\usepackage[utf8]{inputenc} % allow utf-8 input
\usepackage[T1]{fontenc}    % use 8-bit T1 fonts
\usepackage{hyperref}       % hyperlinks
\usepackage{url}            % simple URL typesetting
\usepackage{booktabs}       % professional-quality tables
\usepackage{amsfonts}       % blackboard math symbols
\usepackage{amsmath,amssymb}
\usepackage{nicefrac}       % compact symbols for 1/2, etc.
\usepackage{microtype}      % microtypography
\usepackage{xcolor}         % colors
\usepackage{graphicx}
\usepackage{graphbox}
\usepackage{tikz}
\usepackage{caption}
\usepackage{varwidth}
\usepackage{wrapfig}

\title{Rethinking the modeling of the instrumental response of telescopes with a differentiable optical model}

\author{%
  Tobias Liaudat\thanks{\href{https://tobias-liaudat.github.io}{https://tobias-liaudat.github.io}}, Jean-Luc Starck, Martin Kilbinger, Pierre-Antoine Frugier\\
  AIM, CEA, CNRS \\ 
  Universit\'e Paris-Saclay \\ 
  Universit\'e de Paris \\
  F-91191 Gif-sur-Yvette, France \\
  \texttt{\{tobias.liaudat, jean-luc.starck\}@cea.fr}\\
  \texttt{\{martin.kilbinger, pierre-antoine.frugier\}@cea.fr}
}

\begin{document}

\maketitle

\begin{abstract}
We propose a paradigm shift in the data-driven modeling of the instrumental response field of telescopes.
By adding a differentiable optical forward model into the modeling framework, we change the data-driven modeling space from the pixels to the wavefront. This allows to transfer a great deal of complexity from the instrumental response into the forward model while being able to adapt to the observations, remaining data-driven.
Our framework allows a way forward to building powerful models that are physically motivated, interpretable, and that do not require special calibration data.
We show that for a simplified setting of a space telescope, this framework represents a real performance breakthrough compared to existing data-driven approaches with reconstruction errors decreasing $5$ fold at observation resolution and more than $10$ fold for a $3$x super-resolution. We successfully model chromatic variations of the instrument's response only using noisy broad-band in-focus observations.%\footnote{bla bla}.
\end{abstract}

\section{Introduction}
In astrophysics and cosmology, as in many other fields of physics, the next decade will bring about a new generation of extremely powerful instruments, such as \textit{Euclid} \citep{laureijs2011} or the \textit{Nancy Grace Roman} Space Telescope \citep{wfirst}. A proper modeling of the instrumental response is an absolute prerequisite to the ambitious science goals of these missions.
Current \textit{data-driven} modeling approaches are not able to cope with the stringent error requirements imposed by the new generation of instruments. They can even be orders of magnitude away from the prescribed error budget \cite{schmitz2020}, and they are not able to model chromatic variations (i.e. with wavelength).

In this paper we illustrate how some new technologies brought about by the Deep Learning (DL) revolution can be leveraged to rethink the way we model the instrument response or point spread function (PSF). To this end, we follow the approach in \cite{liaudat2021_b}, and propose a \textit{paradigm shift} in the form of a novel framework that includes a \textit{differentiable optical forward model}. This allows to shift the usual data-driven modeling space from the pixels to the wavefront and translate much of the modeling complexity into the forward model. 
The use of DL methods for the modeling of instrumental response fields is limited due to the complexity of the modeling problem and the fact that current efforts are blind to the physics of the problem \citep{jia2020b}.
The present work lays the groundwork for the introduction of physically-motivated and interpretable DL methods into the modeling of instrumental response fields.

\section{Point spread function modeling for space mission telescopes}
\label{sec:PSF_modeling}
The modeling of the PSF in the field of view (FOV) can be seen as an inverse problem.
Telescope observations include point-like stars and extended objects of interest at different positions in the FOV. We consider stars as samples of the PSF field in the FOV and use them to constrain our model, which we then use to infer the PSF at target positions. 
PSF field modeling encompasses several challenges:
\begin{itemize}
    %
    % Spatial variations 
    \item[1.] \textit{The PSF varies spatially in the FOV}. The model needs to capture the spatial variations of the PSF shape from the stars in order to infer the PSF at target positions.
    %
    % Super resolution
    \item[2.] \textit{The observations are in general under-sampled}. The model needs to super-resolve the output PSFs. This differs from the usual super-resolution (SR) task, as we do not have several observations of the same object we need to super-resolve. In this case, we have several samples of the under-sampled PSF field at different positions in the FOV.
    %
    % Chromatic variations
    \item[3.] \textit{The PSF varies as a function of wavelength}. Also known as chromatic variations, they need to be included in the PSF model for most science goals. 
    When the instrument has a broad passband, each star observation is integrated with respect to its wavelength throughout the passband. 
\end{itemize}
\subsection{Related work}
\paragraph{Data-driven PSF models}
Classical data-driven PSF models only rely on the stars to build the model in pixel space and are blind to the physics of the inverse problem. They mostly differ in the way they handle the spatial variations and the super-resolution \citep{liaudat2020, bertin2011, schmitz2020, jarvis2020, miller2013}. They have difficulties in modeling complex PSF shapes such as those from space missions which are close to the diffraction limit. There is no published method capable of successfully modeling the chromatic variations.
\paragraph{Parametric PSF models}
This class of models, on the contrary, builds a parametric model of the entire optical system that should be as close as possible to the actual telescope. Then, a few number of model parameters are fit to the observations, in some cases wavefronts. Errors will arise when there is a mismatch between the parametric model and the ground truth. Furthermore, even if, ideally, there were no mismatch, the optimization of these models in wavefront space is a degenerate problem. It requires potentially expensive calibration information, usually in the form of out-of-focus observations, in order to break degeneracies. Nevertheless, a model of this type has been used for the Hubble Space Telescope (HST) \citep{krist1993, krist2011}.
\paragraph{Phase retrieval with automatic differentiation}
Estimating the wavefront of an in-focus observation falls in the category of phase retrieval problems. 
Recent works \citep{Wang2020, Wong2021}, based on a framework established by \cite{Jurling2014}, are tackling this problem \citep{schechtman2015} using automatic differentiation. Their objective is solely to estimate the wavefront of a single in-focus image.
Such methods are unusable for the current PSF modeling problem as we are interested in the pixel representation at positions where we do not have observations of the PSF.
In the proposed model, the wavefront is an intermediate product for our goal rather than as an objective in itself.

\section{Data-driven wavefront PSF model}
The stars observed in our FOV are samples of the PSF field. 
The following observational model, 
\begin{equation}
    \bar{I}(x_i, y_i) = \mathcal{F} \left\{ \int_{\text{passband}} \text{SED}(x_i, y_i; \lambda) \; I(x_i, y_i; \lambda) \; \text{d}\lambda \right\} + \mathbf{n}_{i}\;,
    \label{eq:poly_star}
\end{equation} 
relates the star observation $\bar{I}(x_i, y_i) \in \mathbb{R}^{N \times N}$ in the FOV position $(x_i, y_i) \in \mathbb{R}^{2}$ with the objective modeling quantity $I(x_i, y_i; \lambda) \in \mathbb{R}^{M \times M}$, which is at a higher resolution with respect to the observations ($M>N$). The desired instrumental response $I$ is integrated in the instrument's passband weighted by the star's normalized spectral energy distribution $\text{SED}(x_i, y_i; \lambda) \in \mathbb{R}_{+}$, which we consider to be known. Then, it is degraded with the operator $\mathcal{F} : \mathbb{R}^{M \times M} \to  \mathbb{R}^{N \times N}$,  which accounts for down-sampling and could include other types of pixel-level degradation (e.g. sub-pixel shifts, pixel response). The observational noise is modeled by $\mathbf{n}_{i}$, which we assume to be Gaussian for simplicity, i.e. $\mathbf{n}_{i} \sim \mathcal{N}(0, \sigma_{i}^{2} \mathbf{I}_{N})$.
The PSF modeling problem consists in estimating $\{I(x_j, y_j; \lambda)\}_{j=1, \ldots, n_{\text{target}}}$ for all the target positions and wavelengths having as input a set of observations $\{\bar{I}(x_i, y_i)\}_{i=1, \ldots, n_{\text{obs}}}$.

\begin{figure}
    \centering
    \includegraphics[width=\textwidth]{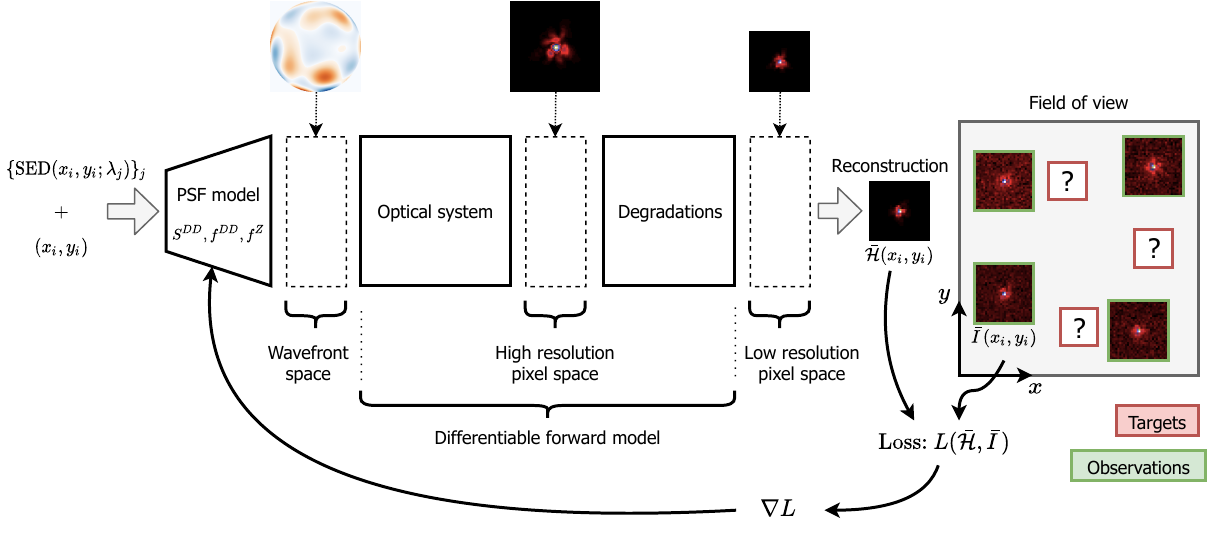}
    \caption{\small A schematic of the proposed framework for data-driven wavefront PSF modeling.}
    \label{fi:psf_model_diagram}
\end{figure}

Up to now, data-driven models were built using some type of dimensionality reduction method applied in \textit{pixel space}, where our observations belong, being blind to the optical system. The modeling problem in pixel space is too complex for the quality and amount of observed data, and they cannot cope with the new challenges of upcoming instruments.
We propose a complete \textit{paradigm shift} with respect to the way data-driven PSF models have been constructed so far. We, instead, model the PSF field in the wavefront space \textit{without any special calibration information}. More specifically, we model the optical system's aberrations as the difference in phase with respect to a wavefront from a flawless optical system. 
This crucial change is allowed by the inclusion of a \textit{physics-based forward model} that can go from the wavefront to the pixel representation. The forward model is end-to-end differentiable thanks to modern automatic differentiation frameworks \citep{tensorflow2015}. It allows propagating gradients from the reconstruction error to the wavefront parameters of our PSF model through the optical system and the degradations. An illustration of the proposed framework can be seen in \autoref{fi:psf_model_diagram}.

The forward model is built under Fraunhofer's diffraction, which is valid for these astronomical telescopes. This allows to relate the wavefront aberrations to the pixel PSF by means of the squared absolute value of the fast Fourier transform (\texttt{FFT}) of the complex pupil function \citep{born1964, goodman2005}. The sampling can be controlled by a variable zero-padding of the pupil function. The reconstructed PSF at any specific wavelength and FOV position can be written as
\begin{equation}
    \hat{\mathcal{H}}(x_i, y_i; \lambda) \propto \underset{p \times p \to M \times M}{\text{crop}} \underbrace{\Bigg| \texttt{FFT}\bigg\{\overbrace{P \odot \exp\Big[ \frac{2 \pi \text{i}}{\lambda} \underbrace{\Phi(x_i, y_i)}_{\text{Wavefront PSF model }} \Big]}^{\text{Pupil function}} \bigg\} \Bigg|^{2}}_{\text{Pixel representation}} \;,
    \label{eq:mono_PSF_model}
\end{equation}
where $P \in \mathbb{R}^{p \times p}$ represents the telescope's obscurations, and $\Phi : \mathbb{R}^{2} \to \mathbb{R}^{p \times p}$ is the wavefront PSF model that maps a position in the FOV with its corresponding wavefront. Once the reconstructed PSF $\hat{\mathcal{H}}$ from \autoref{eq:mono_PSF_model} goes through \autoref{eq:poly_star}, it can be matched with the observations (see \autoref{fi:psf_model_diagram}).

With the proposed framework, a big part of the problem's complexity is shifted from the PSF model into the forward model that is now encoding all the diffraction phenomena. 
The chromatic variation due to diffraction are described by the $\lambda$ dependence in \autoref{eq:mono_PSF_model}.
This allows to simplify the building of the PSF model block seen in \autoref{fi:psf_model_diagram}.
We now need to focus on a good generalization capability of the PSF to target positions, and the model adaptation to the observed data.
We propose to build the wavefront PSF model, $\Phi$, using a weighted sum of wavefront features. The weights will vary as a function of the FOV position $(x,y)$, which writes $\Phi(x,y) = \sum_l f^{\text{Z}}_l(x,y) S^{\text{Z}}_l + \sum_k f^{\text{DD}}_k(x,y) S^{\text{DD}}_k$ with $f^{\text{Z}}_l, f^{\text{DD}}_k : \mathbb{R}^{2} \to \mathbb{R}$ and $S^{\text{Z}}_l, S^{\text{DD}}_{k} \in \mathbb{R}^{p \times p}$. 
The features $S^{\text{Z}}$ are not learned, but based on Zernike polynomials \cite{noll1976}. These provide a good basis for modeling circular apertures as they are orthogonal on the unit disk. 
The data-driven (DD) features $S^{\text{DD}}$ are non-parametric and entirely learned from the data. These DD features adapt to capture variations that are not well modeled by the Zernike features (or modes) and also account for mismatches between our forward model and the ground truth. For $f^{\text{Z}}_l, f^{\text{DD}}_k$ we emphasize on the generalization, and we use for each FOV position polynomials up to orders $d^{Z}$ and $d^{DD}$, respectively. For example, if $d^{\text{Z}}=1$ we have $f^{\text{Z}}_1(x,y) = a_1 + b_1 x + c_1 y$.

In a nutshell, we optimize over the polynomial coefficients of $f^{\text{Z}}_l, f^{\text{DD}}_k$, and the DD features $S^{\text{DD}}$. The loss function used is the mean squared error weighted by the inverse of the estimated noise level, e.g. $L \propto \sum_i (1/\hat{\sigma}_i) \| \bar{I}_i - \bar{\mathcal{H}}_i \|^{2}_{F}$. The optimizer used is the Rectified ADAM \cite{liu2020} algorithm. If prior information is available, it can be added as regularizations at the wavefront level or the pixel level. 

\section{Experiment}
\label{sec:application_euclid}
To demonstrate our novel framework, we simulate a simplified FOV with $2000$ star observations for training and $400$ noiseless target stars for testing. All the positions are randomly distributed in the FOV, and the observations have a variable signal-to-noise-ratio (SNR) uniformly distributed in the range $[10, 110]$. 
We use parameters for an optical model close to the \textit{Euclid}'s VIS instrument model \citep{laureijs2011, venancio2020}, but do not consider further detector effect. For the SEDs, we randomly choose for each star one of the $13$ templates from \citep{pickles1998} as done in \citep[\S5.3]{schmitz2019}. For the ground truth (GT) model we use $45$ Zernike modes with a $d^{\text{Z}}$ of $2$ for each mode. The polynomial coefficients are randomly chosen taking care that the total aberration at any position is within certain limits.
The dimensions used are, $p=256$, $M=64$ and $N=32$.

We compare: \textit{i)} a wavefront PSF model with $15$ Zernike modes and without DD part using $d^{\text{Z}}=2$, \textit{ii)} a model equivalent to \textit{i)} but increasing the Zernike modes to $40$, \textit{iii)} a wavefront PSF model with $15$ Zernike modes, $d^{\text{Z}}=2$, and  $21$ DD features ($d^{\text{DD}}=5$), \textit{iv)} a widely-used model \texttt{PSFEx} \cite{bertin2011}, and \textit{v)} the current state-of-the-art data-driven model \texttt{RCA} \citep{schmitz2020} which has been specially designed for the \textit{Euclid} mission.
All the models are compared in the reconstruction of the target stars to evaluate their generalization power in addition to their modeling capabilities. The metrics analyzed are the root mean squared error (RMSE) of the reconstruction of target stars, at $1$ and $3$ times the observation resolution.

\paragraph{Results}
\begin{wraptable}[12]{r}{8cm}
\centering
\caption{Target star reconstruction RMSE at the observation resolution and at SR.}
  \begin{tabular}{lll}
    \toprule
    &\multicolumn{2}{c}{RMSE [$\times 10^{-5}$] (relative)}   \\
    \cmidrule(r){2-3}
    PSF model   & Resolution x1 & Resolution x3 \\
    \midrule
    i) Zernike 15                   &  $72.3$ ($10.0\%$)    & $18.3$ ($12.4\%$)     \\
    ii) Zernike 40                  &  $22.2$ ($3.0\%$)     & $5.75$ ($3.9\%$)      \\
    \textbf{iii) Zernike 15 + DD}   &  $\mathbf{8.34}$ ($\mathbf{1.1\%}$)     & $\mathbf{4.47}$ ($\mathbf{3.0\%}$)      \\
    iv) \texttt{PSFEx}              &  $69.2$ ($9.5\%$)     & $66.3$ ($43.0\%$)     \\
    v) \texttt{RCA}                 &  $39.6$ ($5.4\%$)     & $85.3$ ($55.5\%$)     \\
    \bottomrule
  \end{tabular}
\label{tb:results}
\end{wraptable}
\setlength\intextsep{0pt}
\autoref{tb:results} summarizes the main results. First, one can notice the \textit{breakthrough in performance} by observing the gap between model \textit{(iii)} and the two models without the forward model (\texttt{RCA, PSFEx}), \textit{almost $5$ times and more than $10$ times lower RMSE for resolutions x$1$ and x$3$}, respectively. One can see that models \textit{(i)} and \textit{(iii)} use a reduced number of Zernike modes with respect to the GT model that uses $45$, three times more. The model \textit{(i)} is under-performing considerably, showing the lack of representation of the reduced number of modes.
However, even if we keep a reduced number of Zernike modes we see the effectiveness of the data-driven features in model \textit{(iii)} and how they can adapt to the observations and still generalize to target positions. 
Despite the increase in the number of Zernike modes in model \textit{(ii)}, being close to the GT, the wavefront DD \textit{(iii)} model still outperforms considerably.
A remarkable fact is the performance gap in the super resolution task between the first three models that use the forward model and the others. It underlines the importance of adding prior physical information into the inverse problem to solve a hard task as SR.
We include the RMSE error as a function of wavelength for model \textit{(iii)} in \autoref{fi:monochromatic_results}, achieving an impressive \textit{mean relative RMSE of $3\%$} showing that the model is capturing the chromatic variations of the PSF. We stress that the other data-driven models, \textit{(iv)} and \textit{(v)}, are not capable of modeling the PSF as a function of wavelength. \autoref{sec:appendix_additional_figures} provides additional figures of this experiment.

\section{Conclusion}
We have presented a novel framework for the data-driven modeling of an instrumental response field which represents a paradigm shift with respect to current state-of-the-art data-driven methods. We propose to include a differentiable physics-based forward model that allows to change the modeling space from pixels to the wavefront. Thus, transferring most of the modeling complexity into the forward model and simplifying the building of the instrumental response model. We have shown the importance of a data-driven term in the model. This term can be successfully learned from noisy under-sampled observations using the differentiable optical model. 
When applied to simplified space telescope simulations, the proposed approach has proven to be a breakthrough in terms of performance compared to current state-of-the-art data-driven models. Furthermore, it is now possible to effectively model the wavelength dependence of the instrumental response from broad-band observations.

\section{Broader impact}
The proposed framework can change the way data-driven instrumental response fields are currently being modeled. This approach can be of particular interest for the \textit{Euclid} space telescope \citep{laureijs2011}, the Vera C. Rubin Observatory \citep{LSST2009} or the \textit{Nancy Grace Roman} space telescope \citep{wfirst}. Weak gravitational lensing\citep{kilbinger2015} plays a fundamental role in their science goals, which places stringent requirements on the point spread function model. Nevertheless, the framework is sufficiently adaptable for other purposes where a good characterization of an instrument response is needed.
If the Fraunhofer approximation is not valid anymore, the optical forward model should be refactored. 
We believe this work does not entail any negative consequences or ethical issues. A journal paper based on this work is in preparation. Code to reproduce the experiences will be released soon.

% %%%%%%%%%%%%%%%%%%%%%%%%%%%%%%%%%%%%%%%%%%%%%%%%%%%%%%%%%%%%

\begin{ack}
The authors would like to thank François Lanusse and Benjamin Remy for useful comments on the manuscript. This work was granted access to the HPC resources of IDRIS under the allocation 2021-AD011011554 and 2021-AD011012983 made by GENCI. This work has made use of the CANDIDE Cluster at the Institut d'Astrophysique de Paris and made possible by grants from the PNCG and the DIM-ACAV.
\end{ack}

\bibliographystyle{unsrtnat}
\bibliography{PSF_prop}

% %%%%%%%%%%%%%%%%%%%%%%%%%%%%%%%%%%%%%%%%%%%%%%%%%%%%%%%%%%%%

\appendix

\section{Additional figures}
\label{sec:appendix_additional_figures}
\begin{figure}[ht]
    \centering
    \includegraphics[width=\textwidth]{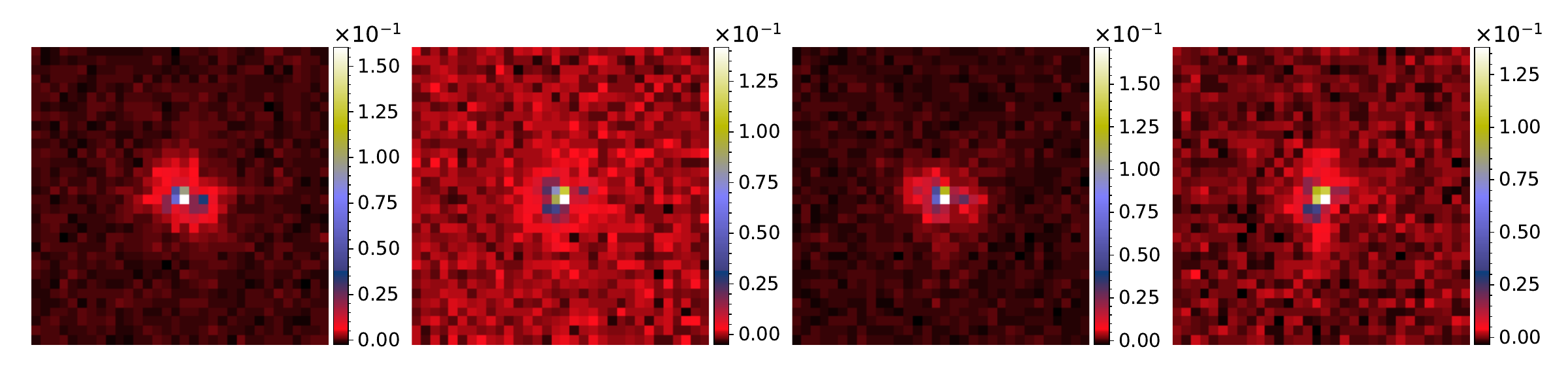}
    \includegraphics[width=\textwidth]{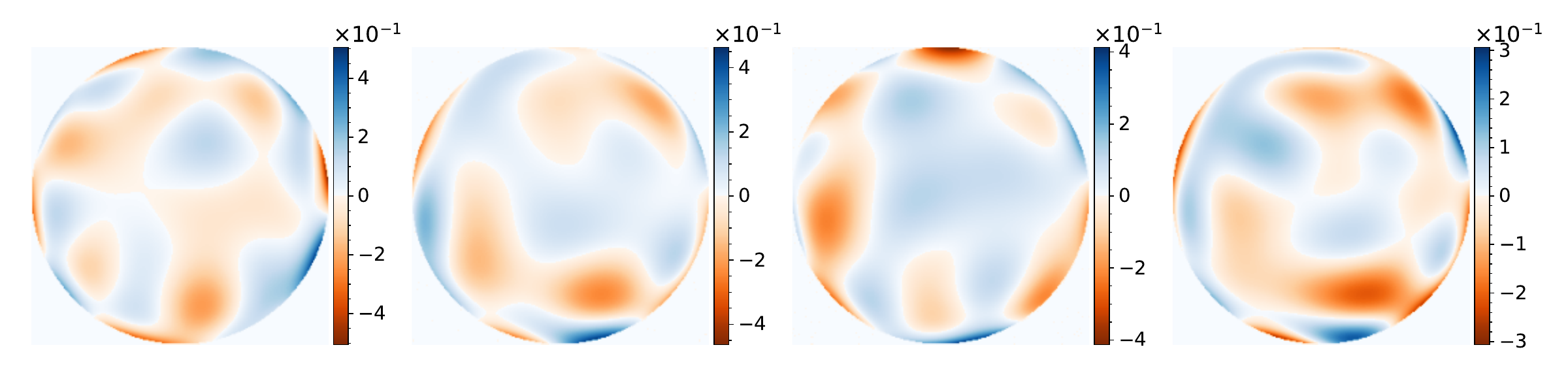}
    \caption{\small The top row presents examples of star observations at \textit{Euclid} resolution from the training dataset. These are examples from different positions in the FOV and have different signal-to-noise ratios. The bottom row shows the corresponding ground truth wavefront maps of each observed star.}
    \label{fi:star_observations}
\end{figure}

In this appendix, we provide additional figures from the numerical experience of \autoref{sec:application_euclid}. \autoref{fi:star_observations} illustrates examples of the observed stars used as training and their corresponding wavefront information.

\begin{figure}
    \centering
    \includegraphics[width=\textwidth]{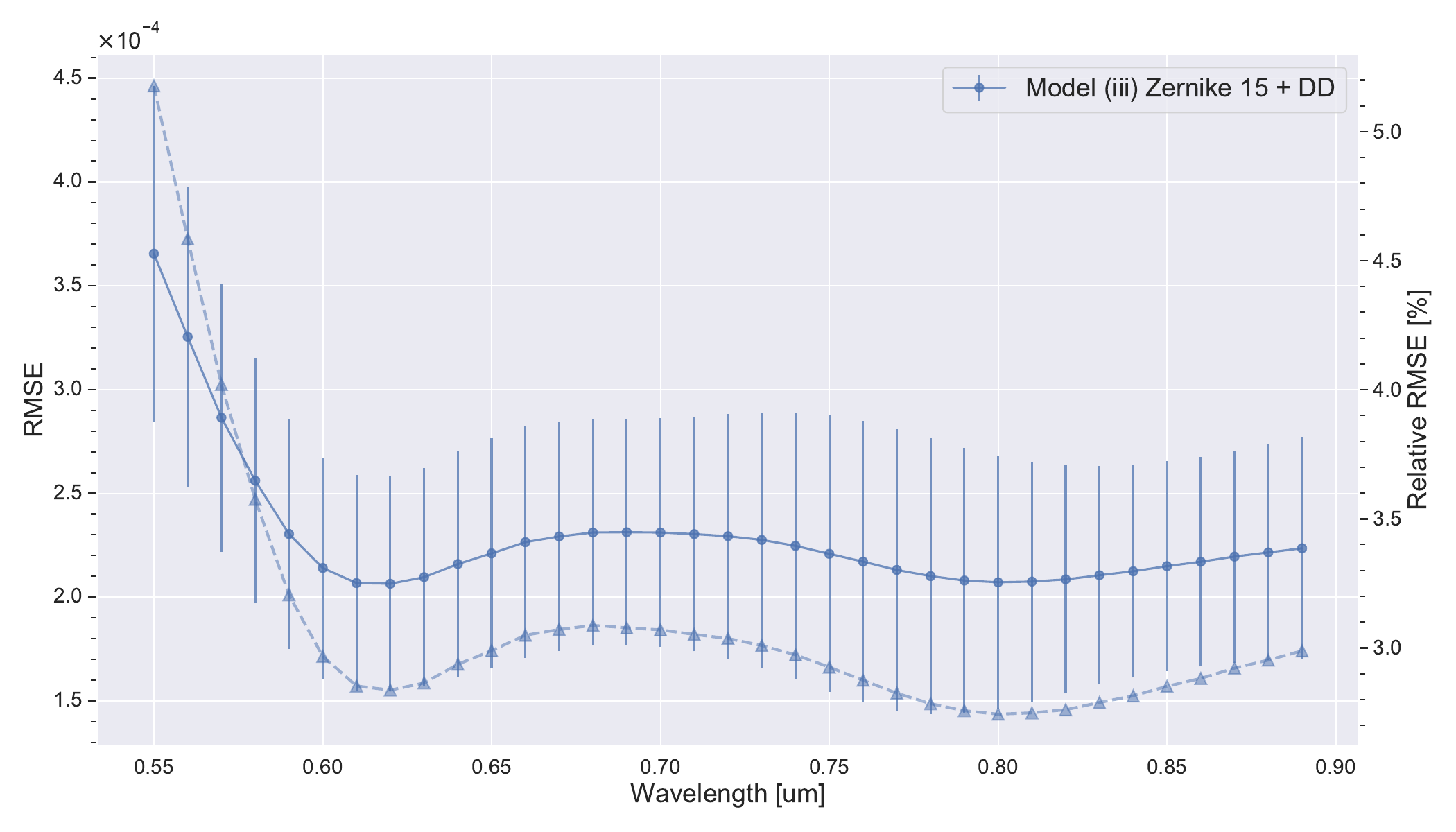}
    \caption{\small Target star reconstruction RMSE as a function of wavelength at three times the \textit{Euclid} resolution. The model \textit{(iii)} uses the proposed wavefront framework with $15$ Zernike modes and a data-driven component. The solid line (circular markers) corresponds to the RMSE on the left axis, and the dashed line (triangular markers) corresponds to the relative RMSE on the right axis. The error bars represent the standard deviation of the RMSE over the star reconstructions at the different target positions.}
    \label{fi:monochromatic_results}
\end{figure}

\autoref{fi:monochromatic_results} presents the target star reconstruction RMSE as a function of wavelength at three times the \textit{Euclid} resolution. We remark that PSF reconstruction at specific wavelengths is not possible for current data-driven PSF models. One can notice that the reconstruction error is kept low inside all of \textit{Euclid's} broad passband (i.e. $550$nm to $900$nm). This indicates that model \textit{(iii)}, in the proposed framework, is not degenerating with respect to the different wavelengths. This could have been the case as the optimization is done only using broad-band observations, as seen in \autoref{eq:poly_star} and \autoref{fi:psf_model_diagram}, without having access to observations at specific wavelengths.

\begin{figure}
    \centering
    \includegraphics[width=\textwidth]{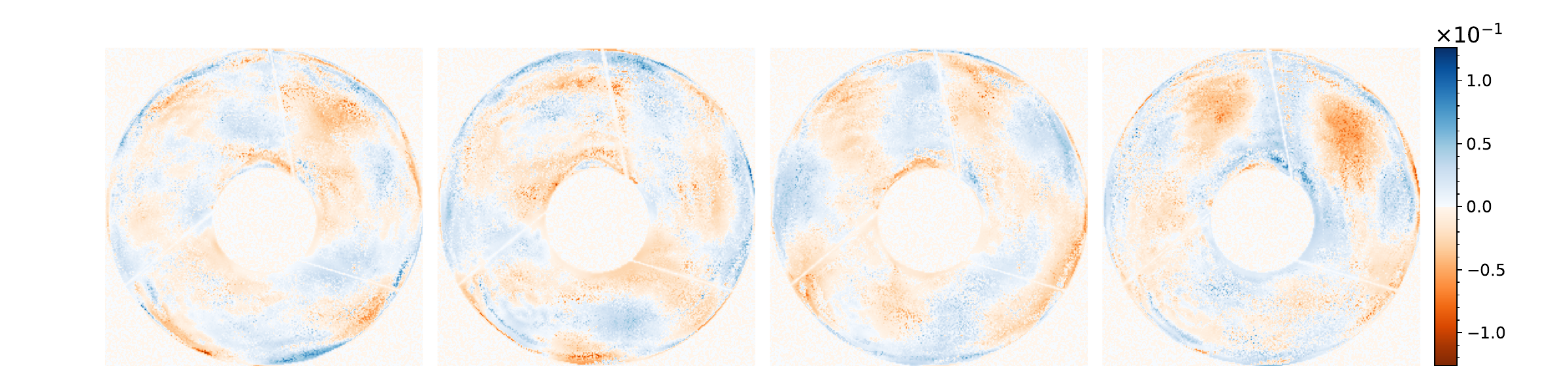}
    \caption{\small Examples of learned data-driven features $S^{DD}$.}
    \label{fi:psf_features}
\end{figure}

\begin{figure}
    \centering
    \includegraphics[width=\textwidth]{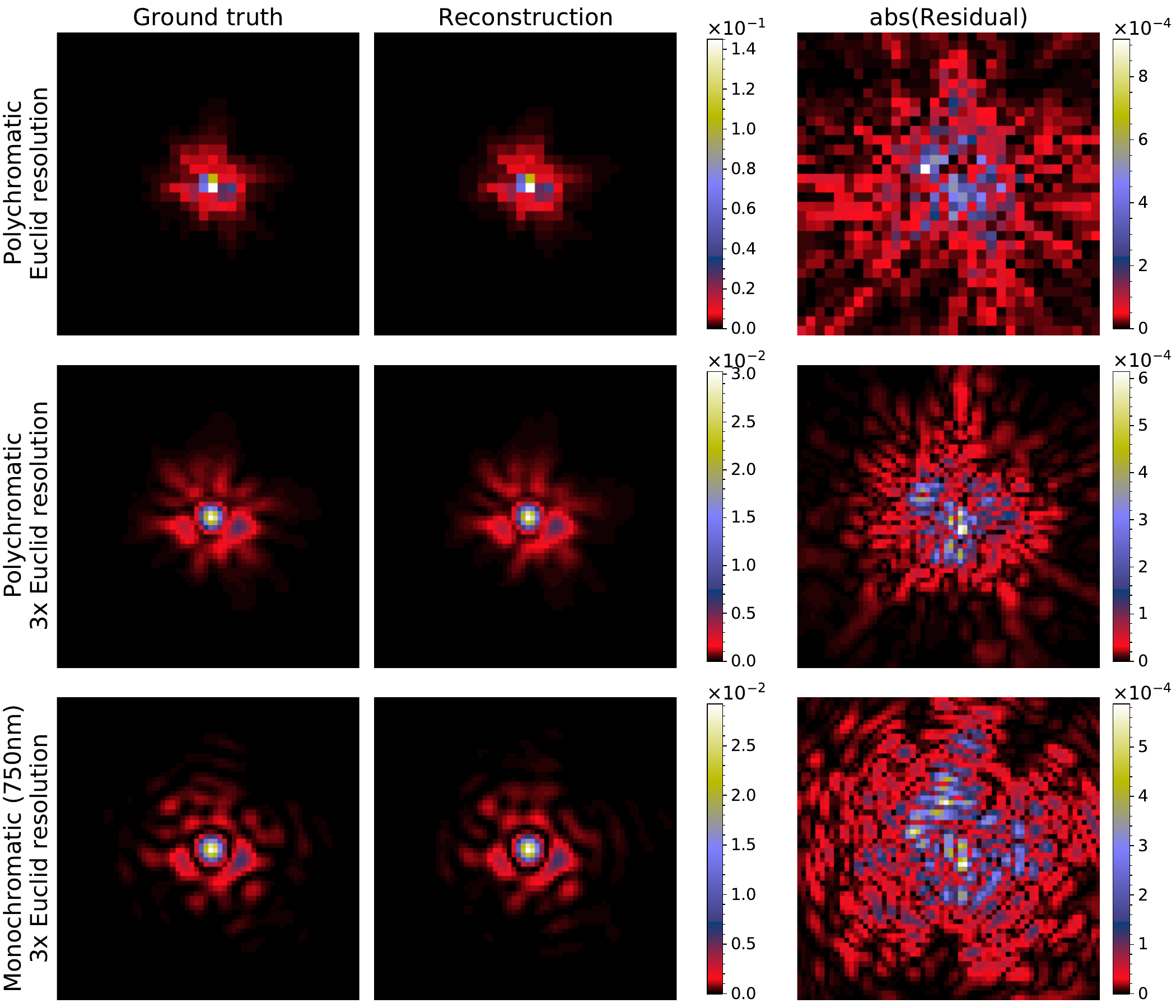}
    \caption{\small Different type of reconstructions for a single target position. The first column represents the ground truth PSF, the second column, the reconstructions from the proposed PSF model using a data-driven component, and, the third column, the absolute value of the residuals between the ground truth and our model. \textbf{Top:} Reconstruction at the \textit{Euclid} resolution for a broad-band, or polychromatic, observation. \textbf{Middle:} Reconstruction at three times the \textit{Euclid} resolution (SR task) for a polychromatic observation. \textbf{Bottom:} Reconstruction at three times the \textit{Euclid} resolution for specific wavelength of $750$nm.}
    \label{fi:psf_reconstructions}
\end{figure}

\autoref{fi:psf_features} shows examples of learned data-driven features from model \textit{(iii)}. It is interesting to see that the learned features are not completely degenerate and some structure has been learned from the stars with the help of the automatic differentiation. \autoref{fi:psf_reconstructions} presents PSF reconstructions at a target position, the corresponding ground truth PSF, and its residuals. We remark the very low residuals obtained by model \textit{(iii)} from the proposed framework. The model is capturing a good super-resolved representation of the PSFs as well as a reliable monochromatic, or single wavelength, representation. It is also proving to generalize well to target positions that have not been used during the training of the model.
One can see the increase of PSF shape complexity when we super-resolve the images as well as when we reconstruct the PSF at specific wavelengths.
These images represent some of the burden in the PSF modeling task and the difficulties we face when trying to estimate the PSF of the third row of \autoref{fi:psf_reconstructions} from observations like the ones from \autoref{fi:star_observations}. It is crucial to change the space of representation of the PSF model from the pixel to the wavefront in order to include the physics of the problem, as for example, the diffraction phenomena.

\end{document}